\documentclass[aps,prb,twocolumn,showpacs]{revtex4}
\usepackage{graphicx} 
\def\gapx{\lower 2pt \hbox{$\buildrel>\over{\scriptstyle{\sim}}$\ }}
\def\lapx{\lower 2pt \hbox{$\buildrel<\over{\scriptstyle{\sim}}$\ }}
\def\he4{$^4$He}
\def\paraH2{{\it p}-H$_2$}
\def\Am2{\AA$^{-2}$}

\begin{document}

\widetext
\title{Melting of a \paraH2 Monolayer on a Lithium Substrate}
\author{Massimo Boninsegni} 
\affiliation{Department of Physics, University of Alberta, Edmonton, Alberta, Canada T6G 2J1}
\date{\today}

\begin{abstract}
Adsorption of \paraH2 films on Alkali metals substrates at low temperature 
is studied theoretically by means of Path Integral Monte Carlo simulations. 
Realistic potentials are utilized to model the interaction between two \paraH2 
molecules, as well as between a \paraH2 molecule and the substrate, assumed 
smooth. 
Results show that adsorption of \paraH2 on a Lithium substrate, the most 
attractive among the Alkali, occurs through completion of successive solid 
adlayers. 
Each layer has a two-dimensional density $\theta_e$ $\approx$ 0.070 \Am2. 
A solid \paraH2 monolayer displays a higher degree of confinement, in the 
direction perpendicular to the substrate, than a monolayer Helium film, 
and has a melting temperature of about 6.5 K. 
The other Alkali substrates are not attractive enough to be wetted by H$_2$ 
at low temperature. No evidence of a possible superfluid phase of  \paraH2 
is seen in these systems.
\end{abstract}
\maketitle
\section{Introduction}\label{intro}
Much theoretical and experimental work has been devoted to the investigation of films of highly quantal fluids (e.g., condensed Helium) adsorbed on a variety of substrates, from the most strongly attractive, such as graphite, to relatively weak ones, such as Alkali metals \cite{bruch97}. The motivation  underlying this effort is the understanding of the fascinating properties that quantum films display, often quite different than those of bulk materials.

Thin films of {\it para}-Hydrogen (\paraH2)  are of considerable interest. One of the reasons is that  a fluid of {\it p}-H$_2$  molecules is regarded  as a potential ``second superfluid", due to the light mass and the bosonic character of its constituents \cite{ginzburg72}. In bulk {\it p}-H$_2$,  superfluidity is not observed, because, unlike Helium,  H$_2$ solidifies at a temperature ($T$ $\approx$ 13 K) significantly higher than that ($\sim$ 4 K) at which phenomena such as Bose Condensation and, possibly, superfluidity (SF) might occur. This is due the depth of the attractive well of the potential between two Hydrogen molecules, significantly greater  than that between two Helium atoms. Several, attempts  have been made   
\cite{bretz81,maris86,maris87,schindler96} to supercool bulk liquid \paraH2, but the search for SF (in the bulk) has so far not met with success.

Confinement, and reduction of dimensionality, are widely regarded as 
plausible avenues to the stabilization of  a liquid phase of \paraH2 at 
temperatures sufficiently low that a SF transition may be observed. Indeed, 
computer simulations yielded evidence of superfluid  behavior  in very small 
(less than 20 molecules) {\it p}-H$_2$ clusters\cite{sindzingre91}, and claims
have been made of its actual experimental observation \cite{grebenev00}.
Theoretically, SF has also been predicted to occur
in a strictly two-dimensional (2d) {\it p}-H$_2$ fluid embedded in a crystalline 
matrix of Potassium atoms \cite{gordillo97}. 

Interesting as the above predictions undoubtedly are, they pertain to 
physical systems that are not easily realized in a laboratory, nor are 
controlled measurements of their properties always practical.
On the other hand, adsorbed films of \paraH2 are readily accessible 
experimentally, and in fact have been extensively investigated, on different 
substrates. For example, the phase diagram and structure of monolayer 
\paraH2 films adsorbed on graphite have been studied by various techniques 
\cite {nielsen80,lauter90,wiechert91,vilches92}. 
One of the most remarkable aspects \cite{vilches92} is that the melting 
temperature $T_m$ of a solid \paraH2 monolayer can be significantly less than 
bulk \paraH2. 
The question arises of whether a considerable reduction of $T_m$ could be 
achieved on a weak substrate, such as that of a Alkali metal. 
It has been shown\cite{myself2} that, for a $^4$He monolayer adsorbed on 
Lithium, zero-point motion in the direction perpendicular to the substrate 
is quite significant, accounting for a doubling of the kinetic energy per 
atom, with respect  to strictly  2d \he4. 
Such zero-point motion, conceivably important in a \paraH2 film as well,  
could result in an effective screening of the interaction of \paraH2 
molecules, leading to a suppression of the melting temperature, perhaps 
to the point where novel quantum many-body phenomena may become observable 
in the liquid phase. 

In this work, attention is directed to the physics of \paraH2 films 
adsorbed on Alkali metals substrates. Besides the issue of SF, this 
system has elicited interest among condensed matter scientists for 
its intriguing wetting properties.  It was suggested \cite{cheng93b}, 
and experimentally verified \cite{cheng93b, mistura94, ross98}, that 
H$_2$, like \he4, ought not wet Cesium (Cs) or Rubidium (Rb) substrates 
at low temperature, due to the relatively shallow depth of the 
substrate-adatom  potential. It is not known whether a thermodynamically 
stable \paraH2 film should exist, at low $T$, on {\it any} of the other 
Alkali substrates, namely Potassium (K), Sodium (Na) or Lithium (Li). 
Particularly intriguing is the question of the existence of  a \paraH2 
{ monolayer} on these substrates, and what its melting temperature $T_m$ 
should be.

Aside from the  study of the phase diagram of 2d \paraH2 by Gordillo and Ceperley \cite{gordillo97}, and of a \paraH2 surface by Wagner and Ceperley \cite{wagner94,wagner96}, 
theoretical calculations for realistic models of adsorbed \paraH2 films have focused on graphite\cite{nho02}. Only recently have the first theoretical studies of \paraH2 films on Alkali substrates been carried out, based on Path Integral Monte Carlo (PIMC) simulations \cite{shi03,szybisz04} as well as on semi-empirical Density Functional methods \cite{szybisz04}. PIMC studies have shown evidence of a wetting transition for liquid \paraH2 on Rb and Cs at temperatures well above 20 K.

Here, we present detailed PIMC results for a model of \paraH2 film adsorbed on  Li and Na substrates, in the temperature range between 2 and 13 K, i.e., significantly lower than that studied in Ref. \onlinecite{shi03}. 
The most accurate potentials currently available are utilized to describe the interactions among \paraH2 molecules, as well as between the molecules and the substrates, which are regarded as smooth (i.e., corrugation is neglected).

The main results of this study are the following: 
\begin{enumerate}
\item {No stable \paraH2 film forms on a Na substrate, in the $T\to 0$ limit; the same is expected to hold true {\it a fortiori} on a K substrate, which is more weakly attractive than Na.}

\item{Adsorption on a Li substrate occurs through completion of 
successive solid layers. A stable solid \paraH2 monolayer exists, 
whose equilibrium coverage (i.e., 2d density)  is $\theta_e=0.070\pm0.01$ 
\Am2.
This is the same value (within statistical uncertainty) found on graphite 
\cite{nho02}, only slightly greater than the theoretically computed 
equilibrium density of 2d \paraH2 (see Ref. \onlinecite{gordillo97}).}
\item {The spread of a \paraH2 monolayer in the direction perpendicular to the substrate is significantly less than for a \he4 monolayer, and its  melting temperature is $\sim$ 6.5 K, i.e., very close to what has been theoretically predicted\cite{wagner96} for 2d \paraH2.}

\end{enumerate}
The quantitative similarity of the results obtained in this work to those involving different substrates, of greatly varying strength and corrugation, suggests that the basic physics of this system is driven by the interaction among hydrogen molecules, and that the role of the substrate is rather marginal.

Although the PIMC method utilized in this work allows for the sampling of permutations of particles, which is essential in order to reproduce in the simulation any effect due to quantum statistics, permutations are not seen to occur, in the temperature range explored. This is consistent with the conclusion reached in previous work by others \cite{wagner96}. In the crystal phase, permutations are suppressed (as in most solids) by the localization of \paraH2 molecules; at higher temperature, on the other hand, though the crystal melts and molecules are less localized, they also behave more classically, as their thermal wave length decreases. Consistently with permutations not being important, i.e. \paraH2 molecules obeying Boltzmann statistics in the temperature range explored here, no evidence of SF can be seen \cite{note}. 

The remainder of this manuscript is organized as follows: in the next
section, the model Hamiltonian is introduced, and a brief description is provided of the 
most important aspects  of the computational method utilized (PIMC), of which comprehensive reviews exist in the literature;  we then illustrate
our results and outline our conclusions.

\section{Model}

Consistently with other theoretical studies, our system of interest
is modeled as an ensemble of $N$ \paraH2 molecules, regarded as point particles,
moving in the presence of an infinite, smooth planar substrate (positioned
at $z=0$). The quantum-mechanical many-body Hamiltonian is the following:
\begin{equation}\label{one}
\hat H = -{\hbar^2\over 2m}\sum_{i=1}^N \nabla_i^2 + \sum_{i<j} V(r_{ij}) +\sum_{i=1}^N U(z_i)
\end{equation}
The system is enclosed in a vessel shaped as a parallelepid, with periodic boundary conditions in all directions. Here, $m$ is the \paraH2 molecular  mass,
$V$ is the potential describing the interaction between two \paraH2  molecules,
only depending on their relative distance, whereas $U$ is the potential
describing the interaction of a helium atom with the substrate, also
depending only on the distance of the atom from the substrate. The 
Silvera-Goldman potential\cite{silvera78} was chosen to model the interaction 
between two \paraH2 molecules; this potential has been shown to provide an 
acceptable quantitative description of bulk \paraH2 in the condensed  
phase\cite{johnson96}. 

The following potential, proposed by Chizmeshya, Cole, and
Zaremba\cite{chi98} was adopted to describe the interaction of a \paraH2 molecule 
with a smooth substrate (i.e., the $U$ term in (\ref{one}) ) 
\begin{eqnarray}\label{ccz}
U(z)=U_\circ(1+\alpha z)e^{-\alpha z}-
{C_{vdw}\ f_2(\beta(z)(z-z_{vdw}))\over(z-z_{vdw})^3}
\end{eqnarray}
with $f_2(x)=1-e^{-x}(1+x+x^2/2)$ and $\beta(x)=\alpha^2 x/(1+\alpha x)$.
The first term in (\ref{ccz}) represents the Pauli repulsion between the electronic cloud of the molecules and the surface electrons, whereas the
second term expresses the Van der Waals attraction. The values of the parameters $U_\circ$, $\alpha$, $C_{vdw}$
and $z_{vdw}$ used here are the ones supplied in Ref. \onlinecite{chi98}.

The above model is, clearly, highly simplified. By far the most important simplification consists of the neglect of substrate corrugation,
whose role is significant for attractive substrates such as graphite \cite{nho02}, but can be expected to be less important on substrates such
as those of Alkali metals, which are relatively weak.

\section{Methodology}\label{method}

The Path Integral Monte Carlo method is a numerical (Quantum Monte Carlo) technique that allows one to obtain accurate estimates of physical averages for quantum many-body systems at finite temperature. 
The only input of a PIMC calculation is  the many-body Hamiltonian (\ref{one}) (which includes the potential energy functions $U$ and $V$). PIMC yields results that are affected only by a (small) statistical uncertainty, as well as by an error due to the finite size of a many-particle system that can be practically simulated on a computer. Computing facilities commonly available nowadays comfortably afford PIMC computations for systems of several hundred  particles; this size is normally sufficient to obtain fairly accurate estimates of energetics and most structural properties of interest. PIMC simulations have been extensively adopted to study physisorption of Helium on a variety of substrates, including alkali metals\cite{myself2}. 

Because thorough descriptions of the PIMC method can be found in the 
literature \cite{ceperley95}, it will not be reviewed here. We only provide 
some relevant details, namely:
\begin{enumerate}
\item{We have experimented in this work with a high-temperature approximation 
for the many-body density matrix $G(R,R^\prime,\tau)$ which is accurate up to 
order $\tau^4$ (see, for instance, Ref. \onlinecite{voth}). This form is not 
as effective as the pair-product approximation (PPA), more commonly used in 
PIMC simulations \cite{ceperley95} of highly quantal fluids, such as Helium. 
We have empirically observed convergence of the energy estimates with a 
time step $\tau = (1/320)$ K$^{-1}$, almost a three times shorter than in 
comparable PIMC studies based on the PPA \cite{wagner94,nho02}. 
The use of the 
high-temperature density matrix considered here simplifies (in our view) 
the calculation, without rendering it unacceptably inefficient 
computationally. 
Accurate estimates for quantities other than the energy, such as film density 
profiles, can be usually obtained with twice as large a time step.}
\item{A multi-level Metropolis scheme based on the staging algorithm 
\cite{pollock84} with free-particle sampling was used to sample many-particle 
paths through configuration space.
This technique has been shown to be an effective remedy to the inefficiency 
of single-slice sampling, at low temperature. We adjusted the length $\bar\tau$
of the portion of individual single-particle paths that the algorithm attempts 
to update, so that the acceptance rate remain above 20\%. 
Typically, $\bar\tau\approx$ 0.025 K$^{-1}$.
Rigid displacements of single
particle paths were performed as well. The average displacement was chosen to ensure an
acceptance rate of approximately 50\%.} 
\item{We have included in our calculation the sampling of cyclic permutations 
of groups of \paraH2 molecules, based on an algorithm very similar to that 
illustrated in Ref. \onlinecite{ceperley95}. This procedure is not restricted 
to sampling groups of 2, 3 or 4 molecules, but allows in principle cycles of 
arbitrary lengths (up to $N$) to be constructed. 
As mentioned in the introduction, however, we found,  at the physical 
conditions explored in this work permutations occur exceedingly rarely. 
This could be the result of using less than optimal a sampling algorithm, 
but we have verified that our procedure is capable of sampling permutations, 
with reasonable efficiency, for bulk $^4$He (a simulated system of 64 atoms) 
at T=2 K and saturated vapor pressure. Thus, we believe the absence of 
permutations in our simulation to reflect a physical effect, in agreement 
with what observed by other authors \cite{wagner94,wagner96}.  }
\end{enumerate}

Physical quantities of interest, besides the kinetic, potential and total ($e$) energy per molecule,
are the \paraH2 density profiles as a function of the distance $z$ from the substrate, i.e. 
\begin{equation} n(z)\equiv {1\over A}\ \int dx\ dy\ \rho(x,y,z)
\end{equation}
where $A$ is the area of the substrate and $\rho(x,y,z)$ is the
three-dimensional (3d) \paraH2 density, as well as the the angularly averaged,
``reduced" pair correlation function $g(r)$, with $r=\sqrt{x^2+y^2}$ and
\begin{equation}
g(x,y) = {1\over  A \theta^2} \
\int dx^\prime dy^\prime \ n (x+x^\prime,y+y^\prime) \ n(x^\prime,y^\prime)
\label{spat}
\end{equation}
where $\theta=N/A$ is the coverage and with $n(x,y)=\int dz \ \rho(x,y,z)$. The correlation function $g(r)$  provides a quantitative assessment of the two-dimensional character of an adsorbed
film; the more two-dimensional an adsorbed film, the more closely $g(r)$
mimics the pair correlation function of a strictly two-dimensional system
of the same coverage.

As for the typical size of the physical systems simulated in this work, for   $\theta \le \theta_\circ=0.072$ \AA$^{-2}$, PIMC calculations were carried out on a system of 36 \paraH2 molecules, initially
arranged on a triangular lattice at a distance of 3 \AA\ from the
substrate. At higher coverage, the initial arrangement was taken to be a series of successive
solid (triangular) layers, arranged as in a hexagonal close-packed lattice, each layer of 2d density $\theta_\circ$ and with
an incomplete top layer. Films of up to three layers are studied. At the monolayer equilibrium coverage $\theta_e=0.070$ \AA$^{-2}$, we also obtained results for a system of 144 particles. No noticeable dependence of the estimates on the size of the system can be observed for the physical quantities discussed here, with the exception of the energy.
In all calculations, the height of the simulation
box (i.e. the distance between the periodically replicated images of the
substrate) is 30 \AA, i.e., much greater than the maximum film thickness observed, so as to make the use of periodic boundary conditions in the
$z$ direction (perpendicular to the substrate) uninfluent. 
\section{Results}
PIMC results were obtained in whis work for two substrates, namely Na and Li. Let us begin with an assessment of the energetics of the \paraH2 films on the two substrates.
\begin{figure}[h]
\centerline{\includegraphics[height=3in,angle=-90]{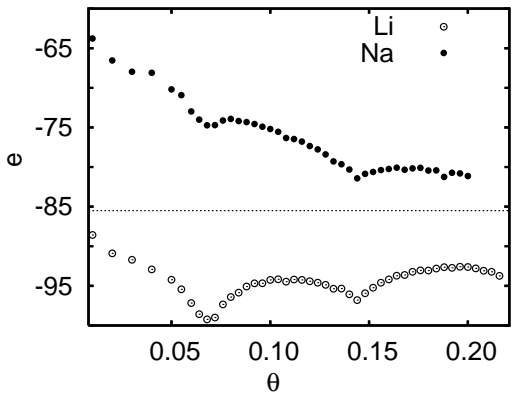}}
\caption{Energy per  molecule  $e$ (in K)  vs. coverage $\theta$  (\AA$^{-2}$), for a \paraH2 film on Li (open symbols) and Na (filled symbols) substrates, computed by PIMC at a temperature $T$=2 K. Dotted line represents the chemical potential of bulk \paraH2 at $T=0$ (from Ref. \onlinecite{silvera}). Statistical errors on the PIMC data are of the order of the symbol size.} 
\label{fig1}
\end{figure}

Fig. \ref{fig1} shows computed values of the energy per \paraH2 molecule $e(\theta)$, as a function of coverage, on a Li and on a Na substrate, at a temperature $T$=2 K. These numbers can be very reliably regarded as ground state estimates, as values of the energy computed below $T\approx$ 5 K are found to be temperature-independent, within the statistical uncertainties of the calculation. The dotted line in Fig. \ref{fig1} represents the ground state energy per molecule ($e_b=-85.5$ K) of bulk \paraH2, in the fcc crystal structure, as computed by Silvera and Goldman \cite{silvera} using self-consistent phonon theory \cite{notex}. 

The two curves shown in Fig. \ref{fig1}  share some qualitative features, e,g.,
a succession of well-defined, equally spaced local minima, corresponding to 
adlayer completion. 
These minima occur at the same values of $\theta$ on both substrates, 
suggesting that the main player in the energy balance is the interaction 
potential among \paraH2 molecules. 

 Let us consider the case of Na first. Although it is (almost monotonically) decreasing, the $e(\theta)$ curve on Na appears to be flattening off, approaching the value $e_b$ from above, asymptotically. This indicates that no film of any thickness will form, i.e., that \paraH2 will not wet Na, at low $T$. Clearly, this conclusion can only be made tentatively, based on the results obtained up to a value of the coverage $\theta=0.2$ \AA$^{-2}$. It is possible that a stable (thick) film may exist. On performing PIMC calculation on systems twice the size of that considered here, one could establish more robustly the trend observed here; however, even in that case one would 
need to supplement the information provided by PIMC with some other calculation (perhaps DF), capable (at least in principle) of making predictions in the $\theta\to\infty$ limit.

Assuming that on Na one indeed has a non-wetting situation, one question that arises is that of the temperature associated to the wetting transition; this issue was not addressed in this work.

On Li, on the other hand, the  $e(\theta)$ curve is monotonically decreasing for $\theta \le \theta_e$, with $\theta_e\approx 0.070$ \AA$^{-2}$; it attains a minimum at $\theta_e$, with $e(\theta_e)\approx-100$ K, i.e. below $e_b$, and displays a slowly increasing trend (while going through well-defined local minima, as mentioned above) for greater values of $\theta$. It seems physically reasonable to expect that $e(\theta)$ should also flatten off, and approach $e_b$ from below, in the $\theta\to\infty$ limit, though, again, the finite size limitation of this calculation do not allow us to make this conclusion firmly. 

The local minima of $e(\theta)$ correspond to successive solid layers (more about this later in the manuscript);  between these minima, films are thermodynamically unstable. This situation is physically quite different from the case of a $^4$He film on the same substrate, for which continuous growth of film thickness as a function of chemical potential is observed, with no layering \cite{myself3}.

\subsection{\paraH2 monolayer on Li substrate}
In the remainder of this manuscript, we shall concentrate on the substrate on which thermodynamically stable \paraH2 films are seen to form, i.e., Li.  Completion of the first layer occurs at a coverage $\theta=0.070\pm0.01$. This value is consistent with that found on graphite\cite{nho02}, even though a graphite substrate is over four times more attractive, and  corrugation plays a significant role. It is also only slightly above the equilibrium density \cite{gordillo97} of pure 2d \paraH2. 
\begin{figure}[h]
\centerline{\includegraphics[height=3in,angle=-90]{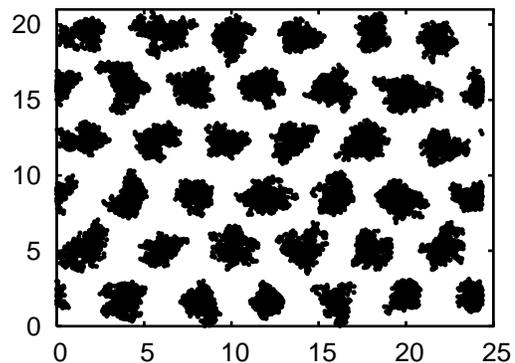}}
%
\caption{Typical many-particle configuration for a \paraH2 monolayer film adsorbed on a Li substrate (top view) at a temperature $T$=2 K. Each black  ``cloud"  represents one of 36 \paraH2 molecules. Periodic boundary conditions are used in both directions. The arrangement of the molecules on a triangular lattice is clearly seen, as well as the absence of overlap among different ``clouds", to show that quantum exchanges are unimportant in this system, at this temperature.}  
\label{fig2}
\end{figure}

This suggests that, as long as the substrate is sufficiently strong to stabilize a film,  the basic physics of an adsorbed \paraH2 monolayer film is largely substrate-independent, driven primarily by the interaction potential between \paraH2 molecules. Fig. \ref{fig2} shows an instantaneous ``snapshot" of a many-particle configuration for an adsorbed film at the equilibrium density (top view). Each fuzzy ``cloud" represents a single molecule; the characteristic size of a cloud gives a quantitative measure of the effect of zero point motion. The arrangement of all molecules on a triangular lattice is fairly clear. Also apparent is the lack of any substantial overlap between different clouds, to indicate that permutations of particles are unimportant in this system.

\begin{figure}[h]
\centerline{\includegraphics[height=3in,angle=-90]{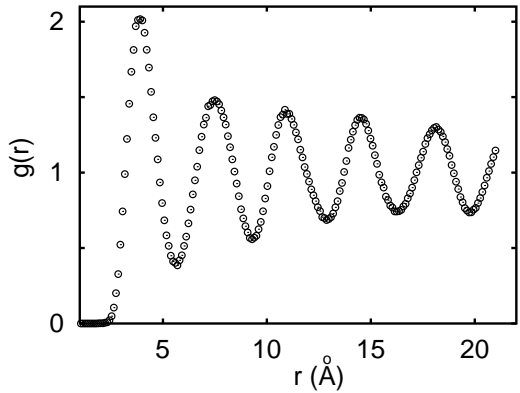}}
%
\caption{Reduced pair  correlation function $g(r)$, defined through Eq.
(\ref{spat}), computed by PIMC for a \paraH2 monolayer film adsorbed on a Li substrate,
at coverages $\theta = 0.070$ \AA$^{-2}$ and $T=2$ K. Statistical errors are smaller than 
symbol sizes. The curve is virtually indistinguishable from the pair correlation function for 2d \paraH2 at the same coverage, at $T$=0.}
\label{fig3}
\end{figure}

More quantitative structural information is furnished by  the density correlation function $g(r)$, defined in Eq. (\ref{spat}) and shown in Fig. \ref{fig3}.   The computed $g(r)$ at $\theta=\theta_e$ is found to be virtually indistinguishable  from the pair correlation function in strictly two-dimensional \paraH2, at the same coverage (and at $T$=0) \cite{myselfun}. 

\begin{figure}
\centerline{\includegraphics[height=3in,angle=-90]{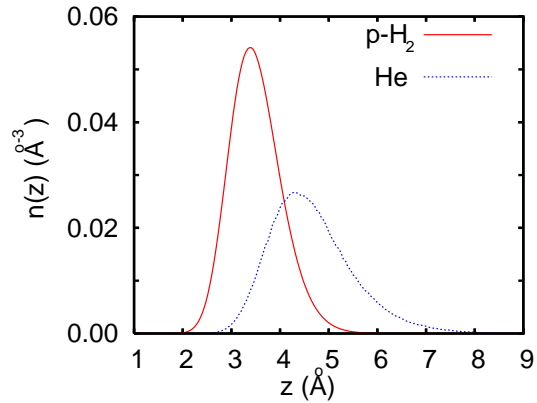}}
%
\caption{Density profile $n(z)$ (\AA$^{-3}$) in the direction
$z$(\AA) perpendicular to a Li substrate, for a \paraH2 monolayer film of coverage $\theta_e=0.070$ \Am2 and at $T$=2 K (solid line). Statistical errors cannot be seen on the scale of the figure. Dotted line shows, for comparison, the corresponding density profile for a monolayer $^4$He film on the same substrate, at $T$=0 and at the equilibrium coverage $\theta_e=0.052$ \AA$^{-2}$ (from Ref. \onlinecite{myself3}). }
\label{fig4}
\end{figure}

Fig. \ref{fig4} shows the \paraH2 density profile in the direction ($z$) 
perpendicular to the substrate (solid line). Also shown, for comparison, 
is the corresponding density profile for a monolayer $^4$He film adsorbed 
on the same substrate, at $T$=0 (from Ref. \onlinecite{myself3}). The \paraH2 
film rests  in closer proximity to the substrate (by approximately 1.5 \AA) 
with respect to the Helium one, and has about half the width. Thus, the 
physics of a \paraH2 monolayer is even more 2d than that of one of He. 
Moreover, the kinetic energy per \paraH2 molecule at this coverage, 
in the $T\to 0$ limit is worth approximately 40 K, i.e., $\sim$ 
30\% more than in a
purely 2d \paraH2 system at the same coverage \cite{myselfun}; this should be 
compared to the factor two difference in kinetic energy between 2d \he4 and 
an adsorbed \he4 film on Li (from Ref. \onlinecite{myself3}).

\begin{figure}
\centerline{\includegraphics[height=3in,angle=-90]{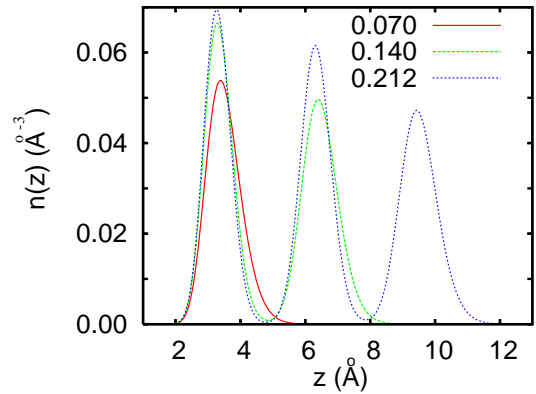}}
%
\caption{Density profiles $n(z)$ (\AA$^{-3}$) in the direction
$z$(\AA) perpendicular to a Li substrate, for \paraH2 films of varying coverage, corresponding to one, two and three layers,  at $T$=2 K. Statistical errors cannot be seen on the scale of the figure. } 
\label{fig5}
\end{figure}

In spite of their lighter mass, \paraH2 molecules experience a lesser degree 
of zero-point delocalization in the perpendicular direction, because of the 
much stronger attraction to the substrate\cite{chi98},  as well as to the 
other molecules.  
As more layers are adsorbed on the substrate, zero point molecular motion in the perpendicular direction is suppressed even further in the underlying layers, which become more compressed (see Fig. 
\ref{fig5}).
\subsection{Melting of a \paraH2 monolayer}
\begin{figure}
\centerline{\includegraphics[height=3in,angle=-90]{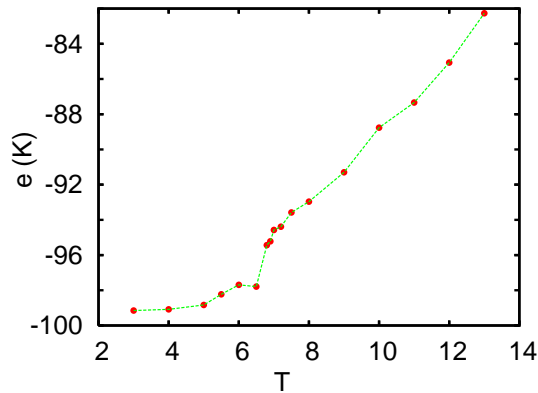}}
%
\caption{Energy per molecule $e(T)$ for a \paraH2 monolayer adsorbed on a Li substrate. The coverage is 0.070 \Am2. The change of slope of the curve at $T\approx$ 6.5 K is evident. Results shown correspond to a system of 144 particles.} 
\label{fig6}
\end{figure}

In order to study the melting of a solid  \paraH2 monolayer at the determined equilibrium coverage (namely $\theta=0.070$ \Am2), we computed the energy per 
molecule of the system as a function of temperature ($e(T)$).   
Melting corresponds to a peak in the specific heat $c_V=(\partial e/\partial T)$. The specific heat can show several such ``anomalies", corresponding to different transitions (e.g., solid to solid-vapour coexistence); several peaks are indeed observed in numerical simulations on graphite, for a commensurate solid layer with vacancies \cite {nho02}.  
In order to reduce the smearing of the peaks, which inevitably occurs when one is performing a numerical study on a system of finite size, we increased the number of particles from 36 to 144 in this calculation, corresponding to a simulation cell in which the area of the substrate $A$ equals 48.74 \AA\ $\times$ 41.21 \AA\ . Results are shown in Fig. \ref{fig6} and summarized in Table \ref{table}. The energy estimates are not corrected for the long-range tail of the interaction potential between \paraH2 molecules, which is truncated at 21 \AA\  in the simulation (potential truncation is routinely done when studying finite systems with periodic boundary conditions). We estimate the correction on the total energy to be worth no more than 0.03 K.

The results of Table \ref{table} show that the dependence of the total energy per molecule on the temperature reflects essentially that of the potential energy alone, as  the kinetic energy depends weakly on temperature, for this adsorbed monolayer film. 
The determination of values for the specific heat from the energy estimates can be made by numerical differentiation, but it is difficult to obtain accurate results in this way. Moreover, our primary objective was to locate the occurrence of the phase transition, and the figure clearly suggests a melting temperature $T_{\rm m}\approx $ 6.5 K, which is where the computed $e(T)$ curve displays a sudden, abrupt change of slope.

\begin{figure}
\centerline{\includegraphics[height=3in,angle=-90]{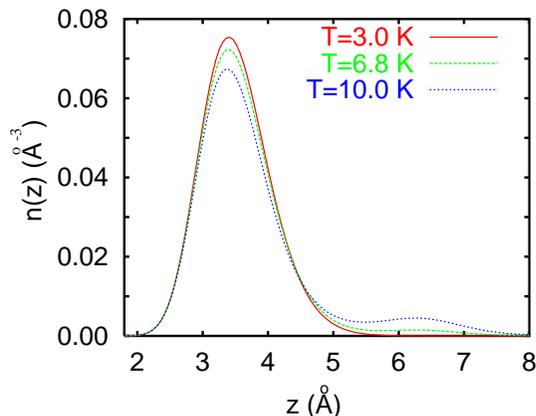}}
%
\caption{Density profiles $n(z)$ (\AA$^{-3}$) in the direction
$z$(\AA) perpendicular to a Li substrate, for \paraH2 films of coverage $\theta=0.070$ \Am2, corresponding to one layer,  at three different temperatures. The melting temperature is estimated at $T_{\rm m}$=6.5 K. Statistical errors cannot be seen on the scale of the figure. As the temperature is raised, second layer promotion of molecules is observed. } 
\label{fig7}
\end{figure}

\begin{table}
\begin{tabular}{|c|c|c|c|}\hline
$T$ & kin & potl & tot \\ \hline
3.0  & 39.86(5)  & -139.007(13)   &-99.15(5) \\
4.0   &39.87(4) &  -138.949(14) &  -99.08(4) \\
5.0   &39.99 (3) &  -138.822(16) &  -98.83(4)\\
5.5   &39.95(7) & -138.183(89)  &  -98.23(6) \\
6.0   &40.07(8)  &-137.77(28)&  -97.69(27)\\
6.5   &40.26(9) & -138.05(11)  &  -97.79(13) \\
6.8   &39.63(14) & -135.06(25) &  -95.43(14)\\
6.9   &39.75(12)&  -134.98(28)&  -95.22(19)\\
7.0   &39.17(15)&  -133.75(20)&  -94.58(14)\\
7.2   &39.40(18)&  -133.79(16)&  -94.39(12)\\
7.5   &39.48(10) & -133.06(17) &  -93.58(12)\\
8.0   &39.04(11) & -132.00(15)& -92.97(16)\\
9.0   &39.16(11)& -130.45(12)&  -91.30(7)\\
10.0  &38.82(8)&  -127.59 (13) & -88.77(10)\\
11.0  &39.43(13)&  -126.76(16)&  -87.33(17)\\
12.0  &40.01(14)&  -125.07(13)& -85.07(13)\\
13.0  &40.01(15)&  -122.27(15)&  -82.27(20)\\ \hline

\end{tabular}
\caption{Kinetic, potential and total energy per \paraH2 molecule (in K) on a Li substrate, computed by PIMC at a coverage $\theta=0.070$ \Am2 at different temperatures. Results are for a 144-particle system. Statistical errors (in parentheses) are on the last digit(s). These estimates are not corrected for the long-range tail of the interaction potential between \paraH2 molecules, which is cut off at 21 \AA. We estimate the correction on the total energy to be worth no more than 0.03 K}
\label{table}
\end{table}

These results are similar to those obtained by Wagner and Ceperley, who studied the melting of a \paraH2 surface \cite{wagner94,wagner96}. The computed melting temperature is also quite close to the theoretical 2d result \cite{wagner96}. This suggests that this system may provide a rather close experimental realization of a 2d system, in many respects even more so than an adsorbed Helium film. 

\begin{figure}[h]
\centerline{\includegraphics[height=3in,angle=-90]{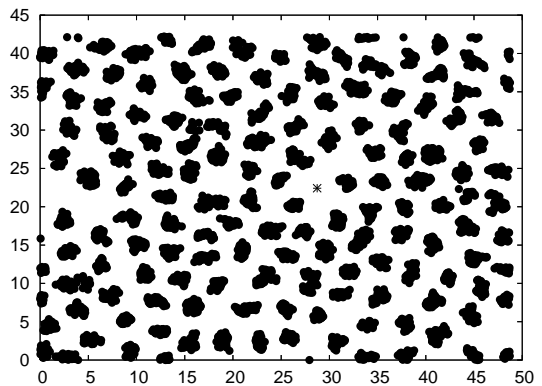}}
%
\caption{Typical many-particle configuration for a \paraH2 film adsorbed on a Li substrate (top view) at a temperature $T$=8 K. Each fuzzy   ``cloud"  represents one of 112 \paraH2 molecules in the bottom (first) adlayer, the remaining 32 molecules in the system having been promoted to the second layer. Periodic boundary conditions are used in both directions. The star at approximately (29,22) indicates the location of a vacancy.}  
\label{fig8}
\end{figure}
The problem of melting of a 2d solid is interesting and, to a considerable degree, still open. In particular, the transition from a solid to a liquid phase can be either discontinuous or continuous, the system going, in the latter case, through an intermediate, ``hexatic" phase featuring local six-fold, bond-angle order \cite{chaikin}. Such a hexatic hase can then melt into a disordered fluid 
as a result of unbinding of dislocations \cite{ktnhy}. 
The results of this study point to a \paraH2 monolayer adsorbed on a Li substrate as a candidate to observe such a scenario experimentally.  

In numerical simulations, on the other hand, unambiguous confirmation of dislocation-mediated melting is complicated by the finite size of the system that can be studied in practice \cite{strandburg}.
Fig. \ref{fig7} shows density profiles computed at three different temperatures, below, above, and close to the melting temperature. They clearly suggest that melting occurs concurrently with second layer promotion; one can therefore opine that melting occurs microscopically with the formation of vacancies in the bottom layer.  Visual inspection of instantaneous many-particle configurations (an example is shown in Fig. \ref{fig8}, for a temperature $T$=8 K) at temperatures just above $T_{\rm m}$  indeed seems to indicate  the presence of vacancies (and dislocations) in the bottom layer.

\section {Conclusions}
Adsorption of \paraH2 on Alkali metals substrates was studied theoretically by means of Path Integral Monte Carlo simulations. In order to assess the expected relevance of the results presented here to experiments, one clearly needs to assess quantitatively the limitations of the model utilized, as well those of the calculation. 

The potential used here to describe the interaction between two \paraH2 molecules, namely the Silvera-Goldman potential,  is  the one for which most of the previous Monte Carlo simulations have been carried out. It is known to be reasonably accurate, even though a comprehensive, quantitative assessment is, to our knowledge, still lacking. Its main limitation is that of being merely a two-body potential. 
Although it is designed to incorporate, at least in part, effects of three-body interactions, this is done only in an effective way. Nonetheless,  it seems unlikely (to us) that  the use of a more refined version of the interaction would significantly, qualitatively  change the results. More
important seems the uncertainty of the substrate-adatom potential. The
potentials used in this calculation are the most current, and are a
significant improvement over the early ``3-9" potential, both with respect
to the functional form as well as the value of the most important parameters
(e.g., the well depth). However, their absolute accuracy has not yet been 
quantitatively assessed; to that aim, additional experimental input may be
needed.

We computed energetics for adsorbed films on both Li and Na substrates; results show that thermodynamically stable adsorbed \paraH2 films will form on Li, whereas a Na substrate (and therefore all other Alkali metals substrates, which are even weaker than Na)  is too weakly attractive for a stable film to form.  This result is consistent with the experimental observation of a wetting transition on Cs and Rb \cite{cheng93b, mistura94,ross98}. Such a transition is predicted to occur on Na and K substrates as well, most likely at a temperature above the critical temperature, as observed on Rb. We also investigated thermodynamic properties of a \paraH2 monolayer on a Li substrate. This system has an equilibrium coverage at low temperature essentially identical to that on graphite, approximately 0.070 \Am2, and is  considerably more two-dimensional that a \he4 monolayer adsorbed on the same (Li) substrate, owing to the much stronger attraction to the substrate experienced by a \paraH2 molecule with respect to a \he4 atom.  We determined the melting temperature of the monolayer to be approximately 6.5 K, which is very close to theoretically computed 2d value. While this value is considerably lower than the 3d one ($\sim$ 13 K), it is still much higher than the temperature at which quantum exchanges are expected\cite{gordillo97} to become significant ($\sim$ 1 K). As exchanges, and long cycles of particle permutations are known to be crucial to the onset of superfluidity \cite{ceperley95},  this system does not appear to offer a promising avenue for the observation of a SF
transition in \paraH2. 
Because a Li substrate is observed to be relatively strong, whereas a Na 
one is already too weak to allow adsorption, one might wonder whether a 
substrate made of an alloy of Li and Na (or, Li and another Alkali metal) 
might have intermediate properties, possibly leading to enhanced zero point 
motion in the perpendicular direction and increased quantum character. 
Theoretical investigation of adsorption on such a ``hybrid" substrate is 
under way.
More generally, however, observing novel phases of molecular \paraH2 may 
require achieving a  substantial renormalization of the interaction of 
\paraH2 molecules, possibly through their interaction with the surface 
electrons of a metal substrate, or of a nanostructure.
\section*{Acknowledgments}
This work was supported in part by the Petroleum Research Fund of the American 
Chemical Society under research grant 36658-AC5, and by the Natural Science 
and Engineering Research Council of Canada under research grant G121210893. 
The author wishes to acknowledge useful discussions with M. W. Cole.


\begin{thebibliography}{99}
\bibitem{notex}
It is worth mentioning that an earlier analysis of experimental data by Schnepp \cite{schnepp70} yielded a lower value, namely $-89.9$ K. In any case, this issue does not alter the main conclusions of this work.
\bibitem{myselfun}
The physical properties of a purely two-dimensional \paraH2 system at 
coverage 0.070 \AA$^{-2}$ in the ground state have been studied separately 
by Quantum Monte Carlo simulations. 
Details of that calculation will be published elsewhere.
\end{thebibliography}
\end{document}